\documentclass[%
superscriptaddress,
 amsmath,amssymb,
 aps,
prb,
floatfix
]{revtex4-2}

\usepackage[utf8]{inputenc}
\usepackage[T1]{fontenc}
\usepackage{amsmath}
\usepackage{hyperref}
\usepackage{natbib}
\usepackage{dcolumn}
\usepackage{bm}
\usepackage{graphicx}
\usepackage{tabularx}
\usepackage{geometry}
\usepackage{indentfirst}
\usepackage{tcolorbox}
\usepackage{float}
\usepackage{multirow}
\usepackage{ulem}
\usepackage{booktabs}
\usepackage[section]{placeins}
\usepackage{hyperref}

\begin{document}

\preprint{Physical Review B}

\title{The role of pressure-induced stacking faults on the magnetic properties of gadolinium }

\author{Rafael Martinho Vieira}
\affiliation{Department of Physics and Astronomy, Uppsala University, Box 516, SE-751 20 Uppsala, Sweden}
\affiliation{Physics, Faculty of Science and Engineering, \AA bo Akademi University, FI-20500 Turku, Finland}
\author{Olle Eriksson}
\affiliation{Department of Physics and Astronomy, Uppsala University, Box 516, SE-751 20 Uppsala, Sweden}
\author{Torbj\"orn Bj\"orkman}
\affiliation{Physics, Faculty of Science and Engineering, \AA bo Akademi University, FI-20500 Turku, Finland}
\author{Ondřej Šipr}
\affiliation{FZU–Institute of Physics of the Czech Academy of Sciences, Cukrovarnická 10, CZ-160 00 Prague, Czech Republic}
\affiliation{New Technologies Research Centre, University of West Bohemia, CZ-301 00 Pilsen, Czech Republic}
\author{Heike C. Herper}
\affiliation{Department of Physics and Astronomy, Uppsala University, Box 516, SE-751 20 Uppsala, Sweden}

\date{\today}

\begin{abstract}
Experimental data show that under pressure, Gd goes through a series of structural transitions $hcp \rightarrow Sm$-type (close-packed rhombohedral)$\rightarrow dhcp$ that is accompanied by a gradual decrease of the Curie temperature and magnetization till the collapse of a finite magnetization close to the $dhcp$ structure. We explore theoretically the pressure-induced changes of the magnetic properties, by describing these structural transitions as the formation of $fcc$ stackings faults. Using this approach, we are able to describe correctly the variation of the Curie temperature with pressure, in contrast to a static structural model using the $hcp$ structure.
  
\end{abstract}

\maketitle
\section{Introduction}

Being the elemental metal with the highest known atomic spin moment at low temperature, Gd seized early the attention of the magnetism community, and its properties are very well documented \cite{Gschneidner1978HandbookEarths,jensen1991rare}. 
Several studies of its magnetic, structural and spectroscopic properties have been reported, from experiments and theory. 
Within the existing literature on first-principles calculations for Gd, it is possible to find different treatments of the $f$-states and their respective impact on the electronic band structure \cite{Locht2016StandardApproximation,Turek2003AbGd}. In conjunction with other studies on magnetic properties such as magnetic moment \cite{Sandratskii1993LocalGadolinium,Sandratskii2014ExchangeStudy}, the exchange parameters \cite{Turek2003AbGd,Locht2016StandardApproximation}, magnetic anisotropy \cite{Colarieti-Tosti2003OriginMetal} and magnetic entropy \cite{Vieira2022RealisticGd}, they have proven to be critical for the description of the fundamental magnetic phenomena and establish a solid background on the capabilities and limitations of ab-initio methods in studying Gd.

Recently, there has been a discussion of the magnetic properties of Gd under pressure~\cite{Li2021NonexistencePressure,Golosova2021HighMetal,Mito2021RelationshipPressures,Mendive-Tapia2017TheoryInteractions}. Experiments report a gradual decrease in the magnetization and the Curie temperature till around 6 GPa, where the magnetization collapses~\cite{Golosova2021HighMetal,Mito2021RelationshipPressures,Samudrala2014StructuralTemperatures,Jackson2005High-pressureTm,Tokita2004RKKYRegions,Iwamoto2003MagneticPressure,McWhan1965EffectHo}, attributing generally such collapse to a shift in the magnetic ordering from ferromagnetic (FM) to antiferromagnetic (AFM). 

Accompanying the changes in the magnetic properties in Gd under pressure, there are also changes in the crystal structure. At around 2 GPa the structure is reported to go from hexagonal-closed-packed ($hcp$) to a $Sm$-type structure ($9R$) and then at 6.5 GPa, close to where the magnetization collapses, there is a structural transformation to a double-hexagonal-close-packed structure ($dhcp$). The coupling between the magnetic and structural properties has not been investigated in detail, which is the motivation behind the present investigation. 

In this work, we show that pure volumetric effects alone do not explain the pressure-induced variations of the magnetic properties. This is in agreement with a previous theoretical study~\cite{Li2021NonexistencePressure}, which found a strong coupling between the stability of the high-pressure structures and the magnetic properties. This work implied that structural changes along the transformation path $hcp\rightarrow Sm$-type ($9R$)$\rightarrow dhcp$ strongly impact the magnetic properties. In this work, we elucidate this finding by means of ab-initio electronic structure theory, coupled with atomistic spin simulations.

The $dhcp$, $hcp$, and $9R$ structures are hexagonal stacked structures that result from a repeating pattern of A, B, and C stacked hexagonal layers with slightly different arrangements of atoms. Depending on the arrangement of these layers, either face-centred cubic, $fcc$, (ABCABC...) or $hcp$ (ABAB...) environments are formed.
For example, the $hcp$ structure (ABAB) contains only sites with $hcp$ environments, while the $9R$ structure (ABABCBCAC) comprises six sites with a $hcp$ environment and three sites with a $fcc$ environment, (see Figure ~\ref{fig:Sm_setup}a). In the $dhcp$ structure (ABAC), there exist two sites with a $hcp$ environment and two others with a $fcc$ environment. 

Since these structures differ primarily in the ratio of $hcp$-$fcc$ environments, one can describe the observed, pressure-induced transformation from $hcp$ to $Sm$-type to the $dhcp$ structure as the formation/accumulation of $fcc$ stacking fault structures \cite{Coles1980TransformationEarths}. This picture is compatible with experiment~\cite{Grosshans1992AtomicAbove}: e.g., the equation of state does not exhibit any drastic variation of the volume or elastic properties at the transition pressures, indicating their second-order nature. Moreover, this perspective can be considered a valuable addition to the existing explanation of coexisting phases expanding or contracting under pressure, as it offers insight into the interface between these phases. The formation and accumulation of periodic stacking fault structures with an $fcc$ arrangement possess a subtle signature structure, making it challenging to differentiate them from systems exhibiting coexisting phases, as detailed in the Appendix~\ref{app:xrpd}.

The appeal of this description of formation/accumulation of stacking faults lies in its ability to allow for a simplified model of the structural changes and reformulates the problem in terms of a relationship between the magnetic properties of Gd and the presence of $fcc$ stacking faults. In the following, we explore this possibility to explain the observed pressure dependence of the magnetization and Curie temperature.

\section{Materials and Methods}
\subsection{Computational details}
Density functional theory (DFT) calculations of the magnetic and electronic structure were made using the RSPt code~\cite{wills2010full}, a full-potential linear muffin-tin orbitals method. Calculations are performed within the local spin density approximation (LSDA)~\cite{Perdew1992AccurateEnergy} and the $f$-electrons are treated as spin-polarized core states (open-core approximation). This setup has been shown to avoid artificial hybridization of $f$-states with $d$-electrons seen in simple DFT calculations of $hcp$ Gd~\cite{Locht2016StandardApproximation}. Moreover, in Ref.~\cite{Li2021NonexistencePressure}, it was observed that the treatment of f-electrons does not play a role in structural stability, so we found it reasonable to use the same setup for the $Sm$-type and the $dhcp$ structures. The package \textit{cif2cell} was used in the preparation of the DFT input files \cite{cif2cell}.

To compare the energies of various magnetic configurations, including those with magnetic disorder in $fcc$ sites, additional calculations were performed using the KKR method~\cite{Korringa1947OnMetal,Kohn1954SolutionLithium} as implemented in SPR-KKR method~\cite{Ebert2011CalculatingApplications}. The magnetic disorder was modelled using the disordered local moments (DLM) approach and treated using the coherent potential approximation (CPA)~\cite{Soven1967Coherent-PotentialAlloys,Stocks1978CompleteAlloys}. For this set of calculations, $f$-states were treated within the LSDA+U correction in the rotationally invariant formulation \cite{Liechtenstein1995Density-functionalInsulators} with an $U$=6.7 eV and $J$=0.7 eV ~\cite{Turek2003AbGd}. 

Using the RSPt software we calculated the Heisenberg exchange couplings, $J_{ij}$, within the LKAG formalism ~\cite{Kvashnin2015MicroscopicFe,Liechtenstein1987}, which we used as input for Monte Carlo (MC) simulations performed within the UppASD code~\cite{Skubic2008AExamples,eriksson2017atomistic} in order to capture the temperature effects of the magnetic properties. In the MC simulations, we made use of the conventional atomistic Heisenberg Hamiltonian: \begin{equation}
  \mathcal{H} = -{1 \over 2} \sum_{i \neq j} J_{ij} \mathbf{e_i} \cdot \mathbf{e_j} - \sum_i \mathbf{H} \cdot \mathbf{m_i}, \label{eq:Heis}
\end{equation} 
which describes the pair exchange interactions between magnetic moments (\textbf{m}) pointing along unit vectors (\textbf{e}), and the interaction of the magnetic moments with an external magnetic field (\textbf{H}). The MC simulations were performed for the $hcp$ as described in Ref.~\cite{Vieira2022RealisticGd}, while for the $dhcp$ and $Sm$-type phases, we considered a simulation box with a minimum of approximately $\approx$83000 sites and 75000 MC steps with the Metropolis algorithm.

\subsection{Model structures}

Whereas the $hcp$ phase of Gd is known to have a ferromagnetic (FM) ordering, the $Sm$-type and $dhcp$ phases have been suggested to be antiferromagnetically (AFM) ordered between pairs of the hexagonal planes (A-type AFM)~\cite{Li2021NonexistencePressure,Golosova2021HighMetal}. Hence, we compared the ground-state energies and magnetic properties of the FM/AFM configurations for the structures under pressure. 

The AFM configuration is incommensurate with the $Sm$-type structure, becoming non-trivial in how the moments in $fcc$-environment sites are aligned. They either become magnetically frustrated ~\cite{Golosova2021HighMetal,Li2021NonexistencePressure} or one has to consider a magnetic unit cell that is twice the structural unit cell, to achieve a lattice that has a proper periodicity of the structure and magnetic properties.

\begin{figure}
  \centering
  \includegraphics[width=0.5\linewidth]{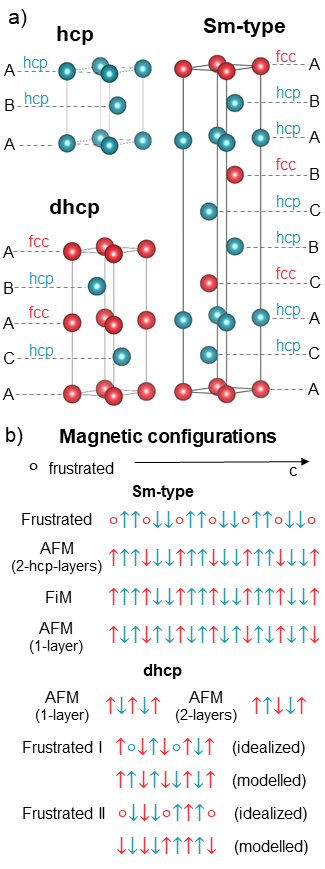}
  \caption{a) Schematic picture of $hcp$ (blue) and $fcc$ (red) environments in the Gd structures considered. Structures generated with VESTA~\cite{vesta}. b)  Diagram of different magnetic configurations considered in the $Sm$-type (top) and $dhcp$ structures.}
  \label{fig:Sm_setup}
\end{figure}

Neutron diffraction data suggests that in the Sm-type structure, the $hcp$ layers form an AFM pattern with two parallel aligned layers (AFM 2-layers), $\uparrow \uparrow$o$\downarrow \downarrow$ intercalated by the $fcc$ layers (o) with frustrated magnetism ~\cite{Golosova2021HighMetal}. With this configuration in mind, we explored the stability of possible AFM arrangements (see Figure ~\ref{fig:Sm_setup}b) in the DFT calculations.
Additionally, we included a ferrimagnetic-like (FiM) configuration, where magnetic moments at the fcc sites have all the same orientation and can be seen as an AFM state with a magnetic defect (caused by the magnetic frustration). 
While not entirely conclusive, the neutron diffraction data in Ref.~\cite{Golosova2021HighMetal} hints at a peak of magnetic origin that can be associated with the (0, 0, 15/2) plane. Within the $dhcp$ structure, the magnetic propagation vector associated with this peak fits the patterns \cite{doi:10.1146/annurev-matsci-070214-021008} presented in Figure~\ref{fig:Sm_setup}b. In the DFT calculations for the $dhcp$ structure, we modelled these configurations, resolving the magnetic ordering in the frustrated sites anti-parallel between them. In this way, we fit the total magnetic moment to zero, in agreement with the experimental observations.

The lattice parameters of the structures considered and magnetic moments resultant from the FP-LMTO calculations at ambient conditions, are summarized in Table \ref{tab:dft_results}. As in other calculations \cite{Li2021NonexistencePressure}, we observe only a tiny variation in the local magnetic moments when different magnetic and structural arrangements are considered. 
In the case of the $Sm$-type structure, we observe in Table~\ref{tab:dft_results} that the AFM (2-layers) state with a doubled structural unit cell has the lowest energy of all arrangements considered in FP-LMTO calculations. This AFM pattern on the $hcp$ layers is consistent with previous findings ~\cite{Golosova2021HighMetal}. However, the very similar energy of the FiM state shows that low-lying, possibly non-collective magnetic excitations are possible for this polymorph. In the KKR-CPA calculation, the ground state is found to be FM instead, but we observe similar energies for the AFM 2-layers ($\Delta E_0$=1.53 mRy/atom) and FiM states($\Delta E_0$=1.59 mRy/atom). We associate the disagreement in the ground state between codes as a result of different approximations and technicalities involved in both methods and accept the results FP-LMTO as a reference due to the general better description obtained in full potential methods. More importantly, we observe that magnetic frustration in the fcc sites (with DLM moments) is not energetically favourable for both FM or AFM configurations. We focus on this last result, as the similarity between the configurations allows for a better comparison of the energies calculated in the KKR-CPA method. This result, combined with the exchange parameters presented ahead, suggests that the magnetic frustration predicted experimentally \cite{Golosova2021HighMetal} does not manifest with the frustrated layers adopting a paramagnetic configuration. Instead, our results suggest that a frustrated layer retains its magnetic order within the layer, but does not align in a periodic pattern along the stacks.

In the case of the $dhcp$ structure, we observe the double-layer AFM configuration to be the ground state, similar to the $Sm$-type structure. The small energy differences between various magnetic configurations hint at a complex magnetic arrangement, with local magnetic excitations.

\begin{table}[h]
\centering
\caption{Summary of the structural and magnetic properties from the DFT (FP-LMTO) calculations at ambient conditions. For AFM configuration different patterns were considered: the 1-layer $\uparrow \downarrow \uparrow \downarrow$ and the 2-layers $\uparrow \uparrow \downarrow \downarrow$ patterns. In the $Sm$-type structure the 2-layer AFM pattern refers only to the $hcp$ layers ($\dagger$). See Figure~\ref{fig:Sm_setup} for a diagram of the magnetic configurations considered.}
\label{tab:dft_results}
\begin{tabular}{ccccc}
\hline
Phase & a (\AA) & c/a & $\langle m_{site} \rangle$  ($\mu_B$) & $\Delta E_0$ (mRy/atom) \\ \hline
\multicolumn{5}{c}{hcp} \\
FM & \multirow{3}{*}{3.642 ~\cite{Vieira2022RealisticGd} } & \multirow{3}{*}{1.60} & 7.59 & 0 \\
AFM (1-layer) &  &  & 7.44 & 3.57 \\
AFM (2-layers) &  &  & 7.53 & 2.31 \\ \hline
\multicolumn{5}{c}{Sm-type} \\
AFM (1-layer) & \multirow{4}{*}{3.618~\cite{tonkov1976conversion}} & \multirow{4}{*}{14.43} & 7.41 & 1.38 \\
AFM $\dagger$ &  &  & 7.49 & 0 \\
FiM $\dagger$ &  &  & 7.49 & 0.01 \\
FM &  &  & 7.53 & 0.31 \\ \hline
\multicolumn{5}{c}{dhcp} \\
FM & \multirow{5}{*}{3.402~\cite{Nakaue1978StudiesDy}} & \multirow{3}{*}{3.25} & 7.41 & 0.25 \\
AFM (1-layer) &  &  & 7.37 & 0.65 \\
AFM (2-layers) &  &  & 7.41 & 0 \\ 
AFM (Frustrated I) &  &  & 7.37 & 0.65 \\
AFM (Frustrated II) &  &  & 7.41 & 0.11 \\ \hline
\end{tabular}
\end{table}

\section{Results and Discussion}

\subsection{Exchange parameters}

\begin{figure}
  \centering
  \includegraphics[width=0.5\linewidth]{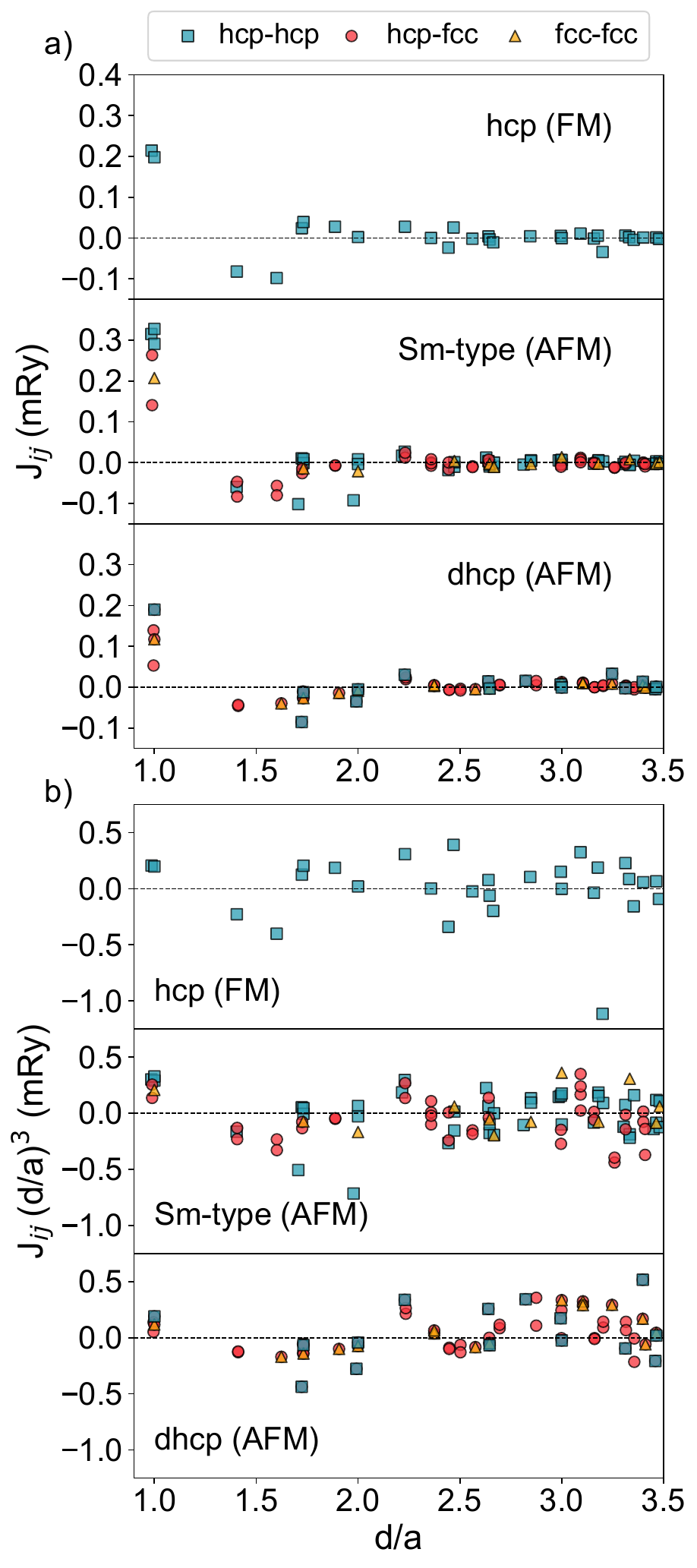}
  \caption{ a) Exchange parameters for Gd in the $hcp$ (top), $Sm$-type (middle) and $dhcp$ (bottom) structures obtained from the respective ground states according to DFT calculations (double layer AFM for both $Sm$-type and $dhcp$). b) Equivalent to a) with a prefactor $(d/a)^3$ on the exchange parameters, to highlight the RKKY interaction. The interaction between different types of sites in the stacking sequence is given as $hcp$-$hcp$, $hcp$-$fcc$ and $fcc$-$fcc$. }
  \label{fig:Jij}
\end{figure}

The exchange parameters ($J_{ij}$) of the three investigated structures ($hcp$, $Sm$-type, $dhcp$) obtained for the respective magnetic DFT ground state configuration are given in Figure~\ref{fig:Jij}. These values serve as input for the subsequent Monte Carlo simulations. The $J_{ij}$ values show similarities between phases, with the most significant disparity being the presence of strong long-range AFM couplings in the region 1.7 $<$ d/a $<$ 2.0 for the $Sm$-type and $dhcp$ phases. Such couplings are likely responsible for de-stabilizing a pure FM order in these structures under pressure. A similar increase in the strength of AFM exchange is observed for elemental Nd when comparing the $dhcp$ phase against the $hcp$ structure~\cite{Kamber2020Self-inducedNeodymium}, which was argued to be responsible for the self-induced spin-glass state of this system.

In general, for the $dhcp$ and $Sm$-type structure, we observe that coupling between first neighbours that appear between an $fcc$ site and an $hcp$ site is weaker than couplings between atoms that both are at $hcp$ sites.

Furthermore, we observe in all phases the oscillatory behaviour of the $J_{ij}$ that characterizes RKKY interactions. The frequency of these oscillations is smaller in the $dhcp$ phase compared to the $Sm$-type and $hcp$ phases (see Figure~\ref{fig:Jij}b). 

\begin{figure}
  \centering
  \includegraphics[width=0.75\linewidth]{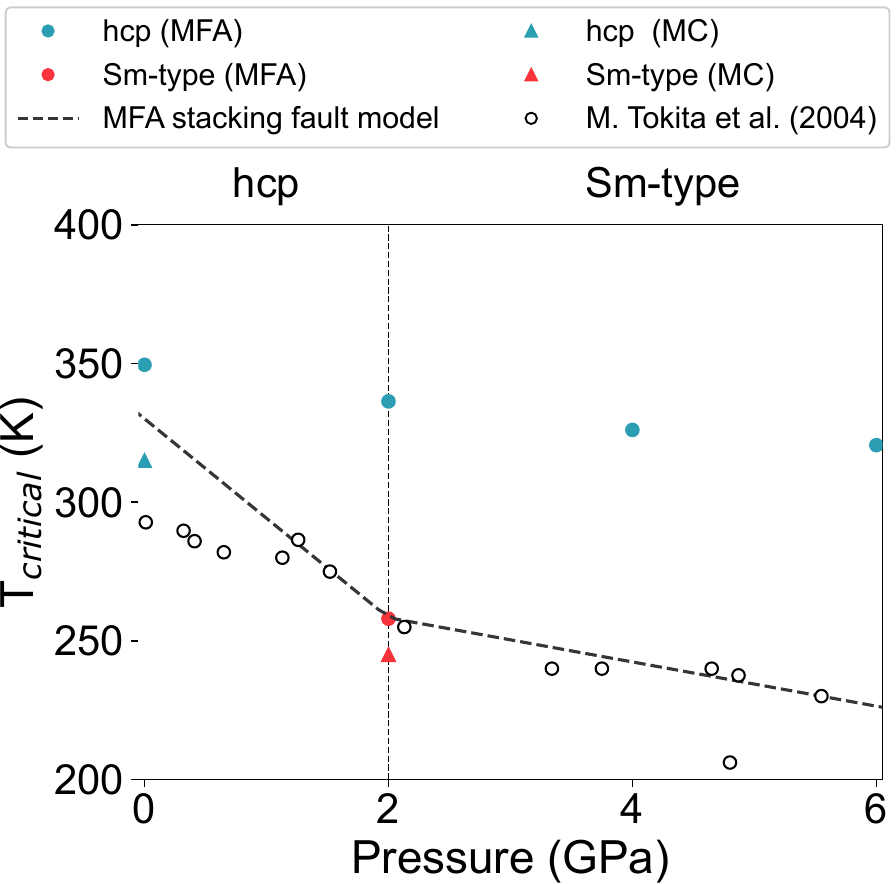}
    \caption{Curie temperature variation with pressure. Open circles refer to experimental data extracted from Ref. \cite{Tokita2004RKKYRegions}, while filled circles are mean-field approximations of $T_{C}$ from the ab-initio exchange parameters. The triangles correspond to the critical temperature calculated from the Monte Carlo simulations. A solid line is used for the mixing model for $hcp$ and $fcc$ stacks to estimate $T_{ C}$ (for more details we refer to the text). }
  \label{fig:TC}
\end{figure}

In Ref.~\cite{Tokita2004RKKYRegions} it is pointed out that the reduction of the lattice parameter by pressure will lower the bottom of the conduction band leading to a decrease in the density of states at the Fermi level ($D(\epsilon_F)$). Since the strength of RKKY interactions is proportional to $D(\epsilon_F)$ ~\cite{Tokita2004RKKYRegions}, this would lead to the decrease of exchange interactions and hence to a lower $T_{C}$. In our first-principles calculations for compressed volumes of the $hcp$ structure (see Table~\ref{tab:hcp_def}), we do indeed observe a reduction of the $D(\epsilon_F)$ that is accompanied by a decrease of the $T_{C}$ (e.g as calculated from mean-field theory, see blue circles in Figure~\ref{fig:TC}). However, as this figure shows, the pressure dependence of the theoretical values is significantly less pronounced than in the experiment.
Additionally, for higher pressures, between 4~GPa and 6~GPa, $D(\epsilon_F)$ increases with the volume reductions, due to intricate details of the electronic structure, while the respective calculation for $T_{C}$ decreases.
This implies that the variation of the RKKY interactions solely by volumetric effects does not explain the observed experimental trend for $T_C$. 

\begin{table}[!h]
\caption{Density of states at the Fermi level vs. pressure (volumetric variations only) in the hcp structure of Gd. The relation between pressure and lattice parameter was determined using the Murnaghan equation of state in the calculations described in Appendix~\ref{app:enthaply}.}
\label{tab:hcp_def}
\centering
\begin{tabular}{@{}cc@{}}
\toprule
Pressure & D($\epsilon_F$) \\
(GPa)          & (mRy/states/atom) \\ \midrule
0              & 12.25           \\
2              & 11.96           \\
4              & 11.44           \\
6              & 11.86           \\
8              & 12.73           \\ \bottomrule
\end{tabular}
\end{table}

An interesting aspect is that we observe a significant increase in the strength of the 1st and 2nd nearest neighbours' exchange interactions (FM) between atoms in $hcp$-like environments for the $Sm$-type structure in comparison to the $hcp$ structure. This result seems counter-intuitive since the experimental trend for $T_C$ with pressure ($dT_C/dP<0$) would suggest that the decrease of $T_C$ is a consequence of the weakening of the ferromagnetic couplings. We observe that instead, an increase in AFM couplings is responsible for the decrease in the ordering temperature, see Figure~\ref{fig:Jij}. Nevertheless, this set of exchange parameters allows us to determine an order-disorder transition temperature close to experimental observations.

\subsection{Monte Carlo simulations}

\begin{figure}
  \centering
  \includegraphics[width=0.55\linewidth]{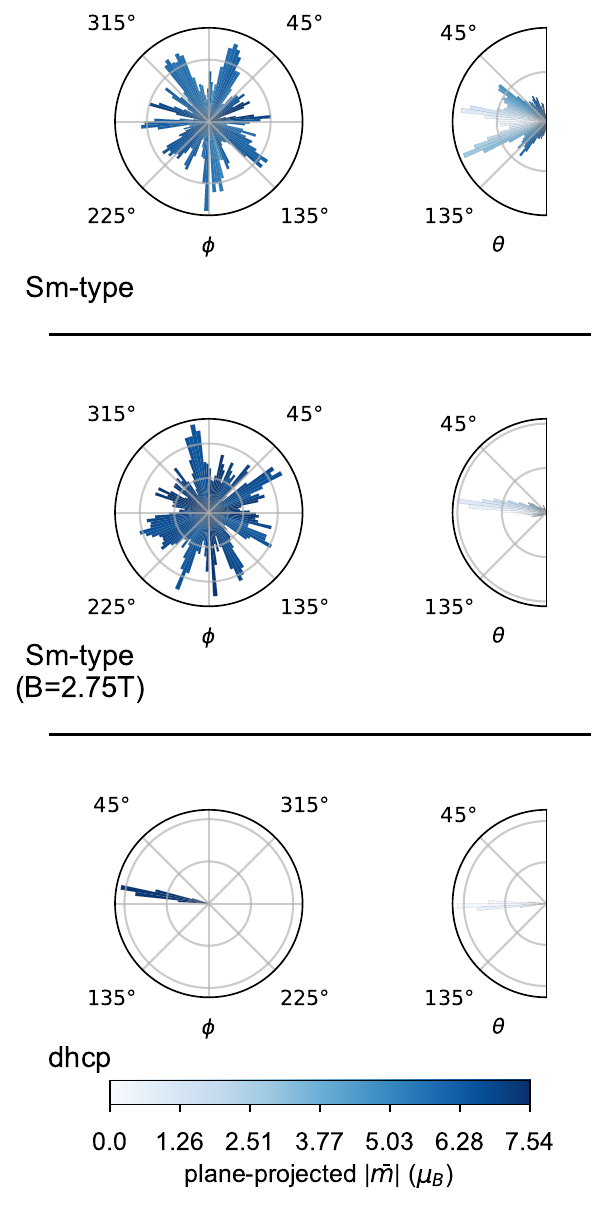}
  \caption{Statistical distribution of spin directions (spherical coordinates with azimuthal angle $\phi$ in the x-y plane and polar angle $\theta$) for different Gd systems simulated in classical Monte Carlo simulations at T=1K with UppASD. The bars are coloured according to the average magnetic moment (projected in the respective x-y or z plane) of the spins aligned in the respective direction.}
  \label{fig:Mag}
\end{figure}

 Materials with comparable FM and AFM couplings tend to develop more complex magnetic configurations, such as spin spirals or spin glasses, in order to balance these competing alignments \cite{Kamber2020Self-inducedNeodymium}. For simplicity, we assume collinear configurations in the DFT calculation. However, the $J_{ij}$ parameters calculated for these configurations should capture this competition between FM-AFM couplings and can therefore be used in MC simulations to get a more realistic picture of the magnetic configuration. 
This analysis is particularly interesting for the $Sm$-type and $dhcp$ structures due to the significant AFM couplings which resulted from these calculations, see Figure ~\ref{fig:Jij}. To examine the relaxed magnetic configuration in the MC simulations, we also determined the distribution of magnetic moment components across the various sites within the simulation box, as illustrated in the histograms presented in Figure~\ref{fig:Mag}.

Experimentally, a finite magnetization is measured for the $Sm$-type phase, which is not in agreement with the AFM configuration predicted by the theory reported in Table 1.  
However, if there is some non-collinearity (e.g. magnetic canting), a finite magnetization can arise in the AFM configuration. A finite magnetic moment can also occur if an external magnetic field is present or due to finite temperature effects which might stabilize a FM configuration over an AFM ordering. Also, if the $hcp$ and $Sm$-type phases coexist, the $Sm$-type phase will experience the magnetization of the $hcp$ phase as an external magnetic field, and hence obtain a finite magnetization. This induced field could, in a very simple model, be included in the interaction between the two magnetic subsystems.

For this reason, we considered two possible setups in the Monte Carlo simulations. First, we simulated a field-free relaxation, guided only by the exchange parameters corresponding to the pure $Sm$-type phase. Afterwards, we proceeded with a MC relaxation under a magnetic field of 2.75T, which is the magnitude of the magnetic field generated by the ordered moments in the $hcp$ phase, to describe the case of coexistence of $hcp$ and $Sm$-type phases, with the latter embedded in the former. Applying an external field is also more realistic for comparison of the calculated magnetization to measurements later. 

\begin{figure}
  \centering
  \includegraphics[width=0.75\linewidth]{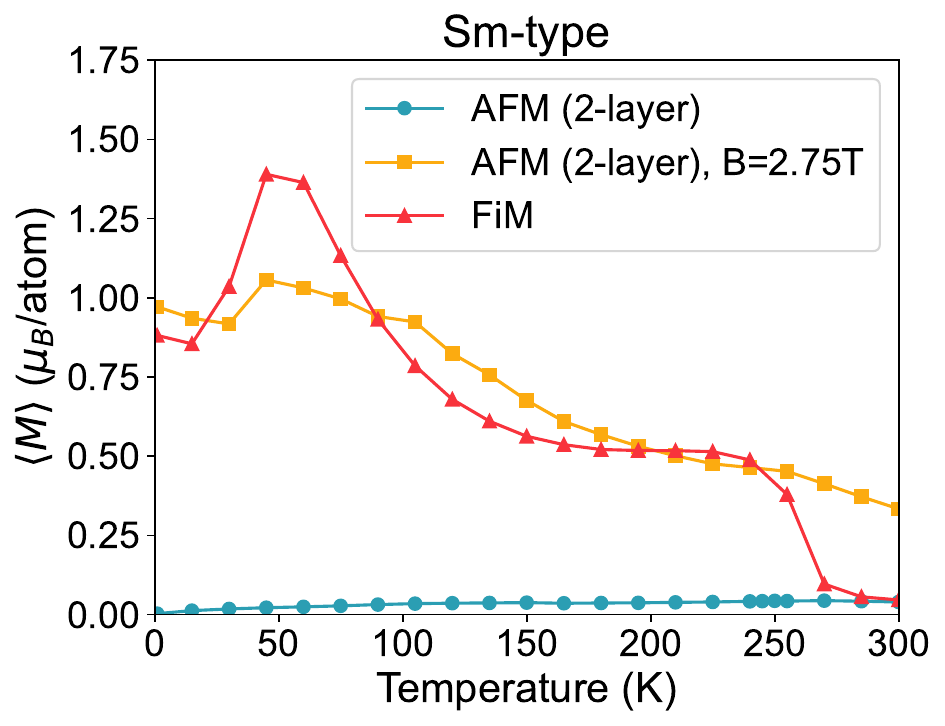}
  \caption{Temperature-dependent magnetization curves simulated with Monte Carlo in UppASD for different sets of $J_{ij}$ of Gd in the Sm-type structure.}
  \label{fig:MC}
\end{figure}

In the field-free simulation, no finite magnetization is obtained in the $Sm$-type phase in the temperature range of 1 to 300~K, see Figure~\ref{fig:MC}. The spin distribution shown in Figure~\ref{fig:Mag}a appears to be roughly an AFM configuration with some degree of disorder but with a clear general orientation of spins. In contrast to that, the setup with applied external magnetic field results in a configuration with a magnetization of roughly 1~$\mu_B$ per atom (Figures ~\ref{fig:Mag}b and ~\ref{fig:MC}). The magnetic field breaks the symmetry, giving rise to a finite induced magnetic moment, while simultaneously increasing the spin disorder in the perpendicular plane, see Fig.~\ref{fig:Mag}b. The calculated heat capacity (see Appendix~\ref{app:heat_capacity}) in the simulation range of 1 to 300 ~K, reveals that both sets of calculations have critical temperatures of around 245 K. Both setups also share a smaller peak on the heat capacity of around 100K. This smaller peak suggests that there is a new magnetic regime likely to be caused by temperature fluctuations overcoming one (or a set of) significant exchange couplings.

A further increase of the pressure leads to the collapse of a magnetically ordered state with finite magnetization and then a transition to the $dhcp$ phase. In this phase, we observed that although the DFT energies for the $dhcp$ phase predict an AFM configuration, the MC simulations converge in a FM configuration with the magnetic moment aligned in the x-y plane, in disagreement with experiments. 
Often, discrepancies between magnetic configurations obtained directly from DFT calculations and configurations obtained from corresponding MC simulations for DFT-derived $J_{ij}$ coupling constants suggest that DFT calculations may not have converged to the ground state. Instead, the magnetic configuration from the MC simulations should be considered for the DFT computations. However, in this case, we confirm that this is not the case, as the energy of the FM configuration is higher than that of the AFM (2-layer) configuration. Additionally, we observe a similar disagreement using the $J_{ij}$ set calculated for the FM configuration, which relaxes into AFM during MC simulations (more details in the Appendix~\ref{app:dhcp}).
Due to the small energy difference between magnetic configurations (see Table~\ref{tab:dft_results}), we infer that the system is magnetically frustrated, potentially due to constraints imposed by the collinear description in DFT or intrinsic properties, similar to what occurs in the case of Nd (dhcp)~\cite{Kamber2020Self-inducedNeodymium}.

\subsection{Stacking faults effects}
In the mean-field approximation (MFA), one can estimate the ordering temperature as 
\begin{equation}
  T_{critical} = \frac{2 J_0}{3 k_B},
   \label{eq:MFA}
\end{equation}
where $J_0 = \sum_j J_{0j}$ corresponds to the sum of the exchange interaction energies \cite{Kvashnin2015}. For structures with stacking faults (such as the $Sm$-type structure), one may identify at least two sets of $J_{ij}$ associated with either $hcp$ or $fcc$ sites. Thus, we can split $J_0$ in Equation~\ref{eq:MFA} in two terms, one containing the sum of all exchange interactions located in $fcc$ sites including then $J_{fcc-fcc}$ and $J_{fcc-hcp}$ couplings), $J^{fcc}_0$, and the sum of all exchange interactions located in $hcp$ sites (including then $J_{hcp-hcp}$ and $J_{hcp-fcc}$ couplings), $J^{hcp}_0$. The two terms are summed, taking into account the fraction of the respective stacking environment, e.g., in the case of the $Sm$-type structure, which is composed of 2/3 of $hcp$ sites and 1/3 $fcc$ sites we use $J_0 = 0.6\overline{6} J^{hcp}_0 + 0.3\overline{3} J^{fcc}_0$, which is then used in Equation~\ref{eq:MFA} to estimate the ordering temperature. 
Note that Equation~\ref{eq:MFA} is suitable for systems with one magnetic sub-lattice, but since we distinguish the $fcc$ and $hcp$ sites having distinct $J_{ij}$ sets we have in practice two sub-lattices. In such cases, the correct procedure for estimation of $T_C$ within the mean-field approximation requires the computation of the highest eigenvalue of the Heisenberg Hamiltonian (see Equation~\ref{eq:Heis}) \cite{UppASDSweden}. Yet, due to their similar properties (magnetic moment and coordination), Equation~\ref{eq:MFA} yields reasonable results for the $Sm$-type phase with only minor discrepancies when compared to the exact approach.

Extending this idea to other stacking structures, we propose to evaluate $J_0$ as an average of both $fcc$ and $hcp$ environment:
\begin{equation}
    J_0 = xJ^{hcp}_{0}+(1-x)J^{fcc}_{0} \label{eq:mean_J}
\end{equation}
with $x$ being the ratio of the number of $hcp$ layers in the structure. This linear interpolation approach aligns with the accumulation model of $fcc$ stacking faults, serving as the transition mechanism between the $hcp$ to $Sm$-type and $dhcp$ structures. While the exact relationship remains to be determined, we have chosen a linear relationship as the simplest assumption. Moreover, the linear relation is also justified by the physical characteristics of the structures considered. Our $J_{ij}$ calculations show that the interaction between layers weakens rapidly with distance (see Figure~\ref{fig:Jij}). Also, we notice that, in general, interactions involving $fcc$ layers are not as strong as pure $hcp$-$hcp$ interactions. Given that there are no adjacent $fcc$ layers in $hcp$, $Sm$-type, and $dhcp$ structures, it becomes clear that $fcc$ layers interact weakly with each other. Thus, the impact of $fcc$ stacking faults on magnetic properties can be approximated to minor perturbations. By combining Equations \ref{eq:mean_J} and \ref{eq:MFA}, we obtain a MFA model for $T_{critical}$ dependence on stacking faults:
\begin{align}
  T_{critical} & = \frac{2 }{3 k_B}\left(xJ^{hcp}_{0}+(1-x)J^{fcc}_{0} \right) \nonumber \\
  & = x T^{hcp}_C + (1-x) T^{fcc}_C  \label{eq:MFA_stacks}
\end{align}
with the magnetic ordering temperature given by a weighted average of the $T_C$ for $hcp$ and $fcc$ layers. We highlight that our approach is analogous to the application of the virtual crystal approximation (VCA)  \cite{PhysRevB.61.7877} to describe alloy behaviour. However, instead of considering chemical mixing, there is mixing of $fcc$ and $hcp$ layers.

To be able to study the pressure dependence of the transition temperature based on 
Equation~\ref{eq:MFA_stacks}, we need a model that relates the fraction of $hcp$ layers, $x$, and $fcc$ layers, $(1-x)$, to the applied pressure. As an approximate description, we consider a linear decrease of $x$ with pressure, with a different decrease ratio between the transition ranges (validity shown in Appendix~\ref{app:enthaply}). Specifically, when transitioning from the $hcp$ phase (where x=1 and is stable at P=0 GPa) to the $Sm$-type phase (where x=2/3 and is stable at P=2 GPa), the decrease ratio has one value, 0.1$\overline{6}$ stacks/GPa. When transitioning from the $Sm$-type phase to the $dhcp$ phase (where x=1/2 and is stable above P=6.5 GPa), a different decrease ratio is applicable, 0.$\overline{037}$ stacks/GPa.

We used this relationship between pressure and the formation of stacking faults, combined with Equation~\ref{eq:MFA_stacks} and the $J^{hcp}_0$ and $J^{fcc}_0$ calculated for the $Sm$-type structure, to evaluate the variation of $T_{critical}$ with pressure. Employing the $J_{ij}$ obtained for the $Sm$-type is motivated by the co-existence of both $hcp$ and $fcc$ sets in the phase and for being roughly the expected structure throughout the majority of the pressure range that was explored (0-6 GPa). 

The calculated $T_{critical}$ values are in good agreement with experimental findings reported in Reference~\cite{Tokita2004RKKYRegions}, as shown in Figure~\ref{fig:TC}.

Here, the agreement in the pressure range of the $Sm$-type structure (2-6.5 GPa) supports the hypothesis of the formation of $fcc$ stacking faults under pressure as the primary mechanism for the $T_{critical}$ variation observed experimentally. The pressure change of the ordering temperature from a model that only considers a volume variation of a stacking-fault-free $hcp$ phase is seen from Figure~\ref{fig:TC} to fail in reproducing the experimental trend, since the so obtained $T_{\rm C}$ barely changes with the pressure (blue circles).

In the region between 0-2 GPa, our model has a less accurate agreement with the experimental data, since it predicts a sharper decrease of $T_{critical}$ compared to the observed value. While the estimated trend aligns with the two experimental data points closest to P=2 GPa, the remaining data appears to be in better agreement with the results based solely on volumetric effects (depicted as blue circles in Figure~\ref{fig:Mag}). Intuitively, it sounds reasonable that the $hcp$ structure would withstand some volume contraction before entering the regime of formation of $fcc$ environments, which would imply a slower variation of $T_{critical}$ when $P \rightarrow 0$ in comparison to the linear approach employed in our model. We also attribute the fast $T_{critical}$ decline predicted at low pressures to the simplification made of not including pressure effects on the exchange parameters. In an ideal case, we expect that $J^{hcp}_0$ and $J^{fcc}_0$ vary in the intermediate structures. This effect has been neglected since we considered the values from the $Sm$-type structure which explains the difference between $T_{critical}$ from the model and the $hcp$ phase at zero pressure. This difference makes it also evident how $J^{hcp}_0$ is more susceptible to the formation of stacking faults than to volume effects (by comparison with $hcp$ in different volumes).

Another consequence of taking fixed $J^{hcp}_0$ and $J^{fcc}_0$ values is the inability to predict the collapse of magnetic ordering near the transition to the $dhcp$ phase. Theoretically, one could improve the model by calculating the $J_{ij}$ parameters of intermediate structures between the $hcp$, the $Sm$-type and the $dhcp$ phases. However, the unit cell needed would increase significantly in the number of atoms, rapidly raising the computational effort for the calculation of the exchange parameters.

In experimental reports, $dT_C/dP$ was determined by linear regression of data points in the whole 0-6 GPa interval being obtained a range 10.6-17.2 K/GPa of values for $dT_C/dP$ \cite{Mito2021RelationshipPressures,Tokita2004RKKYRegions,Iwamoto2003MagneticPressure,Jackson2005High-pressureTm,Golosova2021HighMetal}, i.e., the change in slope beyond 2 GPa was ignored. If the regression is done separately for the 0-2 GPa and 2-6 GPa regions, we obtain distinct slopes for the respective regions (see Appendix~\ref{app:linear_fit}). This behaviour is in line with our hypothesis of the magnetism being critically coupled to the number of stacking faults, which should have different formation rates in these intervals.

Regardless of the mechanism of $hcp\rightarrow Sm$-type$\rightarrow dhcp$ transition description, it implies the presence of at least two phases, either due to spontaneous formation under pressure, growth, or stacking fault accumulation. Since the $hcp$ and the $Sm$-type phase have very distinct magnetization, we tried to ﬁnd the fraction ($\rho$) of $hcp$-phase at each pressure by ﬁtting: 
\begin{equation}
   M_{expt}(P)=\rho M_{hcp}(T) + (1-\rho ) M_{Sm}(T)
\end{equation}
to the magnetization-pressure data from Ref.~\cite{Tokita2004RKKYRegions}. As the temperature associated with the experimental data is unclear, we tried different temperatures in the range between 1-300 K  using the magnetization from our Monte Carlo results. Unfortunately, the results obtained are not comparable with the estimation from volumetric analysis in Ref.~\cite{Golosova2021HighMetal}, which predicts a fraction of $Sm$-type of about 65\% at 2 GPa. In comparison, our closest value was 23\%. Notably, the coexistence of distinct magnetic configurations in the phases ($hcp$, $Sm$-type, and $dhcp$) suggests that their interaction gives rise to an intricate magnetic arrangement. It becomes apparent that the experimental drop in magnetization cannot simply be explained by changes in local moments in the individual phases or simple changes in magnetic ordering. A more sophisticated approach which describes the interaction between the magnetic phases is necessary to accurately describe this phenomenon. In order to do so, measurements of the magnetic order of bulk Gd at different pressures would offer important clues for the theoretical modelling.

\section{Conclusions}
In this work, we investigate the causes for the observed pressure-dependence of the magnetic properties of bulk Gd from first-principles calculations. As reported for Nd, another rare-earth element, we observe that the exchange parameters depend on the stacking structure, with the appearance of additional AFM couplings for the $Sm$-type structure ($9R$) and $dhcp$. 
With this in mind, we describe with success the variation of the magnetic order-disorder temperature along the 0-6 GPa pressure range as linearly dependent on the formation of $fcc$ stackings. By doing so, we identify two different $T_{critical}$/P rates, as a consequence of the different stacking fault formation rates with the pressure associated with the $hcp$ $\rightarrow$ $Sm$-type transition and $Sm$-type $\rightarrow$ $dhcp$ transitions. Although this change in behaviour is not considered in previous reports, the experimental data show a similar shift in pressure dependence. 

Despite the strong coupling between structure and magnetic properties observed, it was possible to describe the pressure dependence of the Curie temperature reasonably well, without considering the pressure-induced effects on the J$_{ij}$ couplings. Moreover, we assume that in the case of phase coexistence, the interaction between phases can be simplified as an external magnetic field created by the respective magnetization. However, this approximation fails to explain the magnetization dependence on pressure and its collapse close to the transition to the $dhcp$ structure. We link this failure to a more complex magnetic configuration resulting from the coexistence of phases with different magnetic ordering which leads to a competition between AFM and FM alignment of the spins.
Our calculations indicate that the $dhcp$ structure may exhibit a complex and frustrated magnetic configuration. Future research, combining theoretical and experimental approaches, could yield valuable insights into the magnetic behaviour of the $dhcp$ phase, potentially uncovering a spin-glass-like state as found in elemental Nd.

\section{Acknowledgments}
RMV would like to thank Anders Bergman for the discussion on magnetic frustrated configurations and Erna K. Delczeg-Czirjak for the support in validation calculations. This work was supported by the Swedish Foundation for Strategic Research, within the project Magnetic Materials for Green Energy Technology (ref. EM16-0039), StandUp, eSSENCE, the Swedish Energy Agency (Energimyndigheten), The Knut and Alice Wallenberg Foundation, the Magnus Ehrnrooth Foundation, the ERC (FASTCORR project) and the Swedish Research Council (VR). The computations performed in this work were enabled by resources provided by the CSC – IT Center for Science, Finland, and by the Swedish National Infrastructure for Computing (SNIC) at NSC and PDC centres, partially funded by the Swedish Research Council through grant agreement no. 2018-05973.

\appendix
\renewcommand\thesection{\Roman{section}}
\section{Theoretical XRPD patterns}
\label{app:xrpd}

To investigate the effect of the accumulation of $fcc$ stacking faults in X-ray powder diffraction (XRPD) patterns, we formulated three theoretical structures with varying ratios of $fcc$ layers between the $hcp$ (0) and Sm-type (1/3) structures. The structures we considered are as in the following patterns ($fcc$ layers in bold): ABA\textbf{B}CBC\textbf{B} (1/4 $fcc$ and 3/4 $hcp$), ABA\textbf{B}CBCBC\textbf{B} (1/5 $fcc$ and 4/5 $hcp$), and ABA\textbf{B}CBCBCBC\textbf{B} (1/6 $fcc$ and 5/6 $hcp$). 

We maintained the lattice parameter $a$ for the $hcp$ phase and adjusted $c$ to match the layer spacing of the $hcp$ structure. Note that the stacking faults are deliberately organized in a periodic manner, making them more detectable in an XRPD pattern. The XPRD pattern of the structures was calculated on VESTA~\cite{vesta} with incident X-rays having a wavelength of 0.68825~\AA, as in Ref.~\cite{Mito2021RelationshipPressures} to ease comparison with experimental data.

\begin{figure}[!h]
  \centering
  \includegraphics[width=\linewidth]{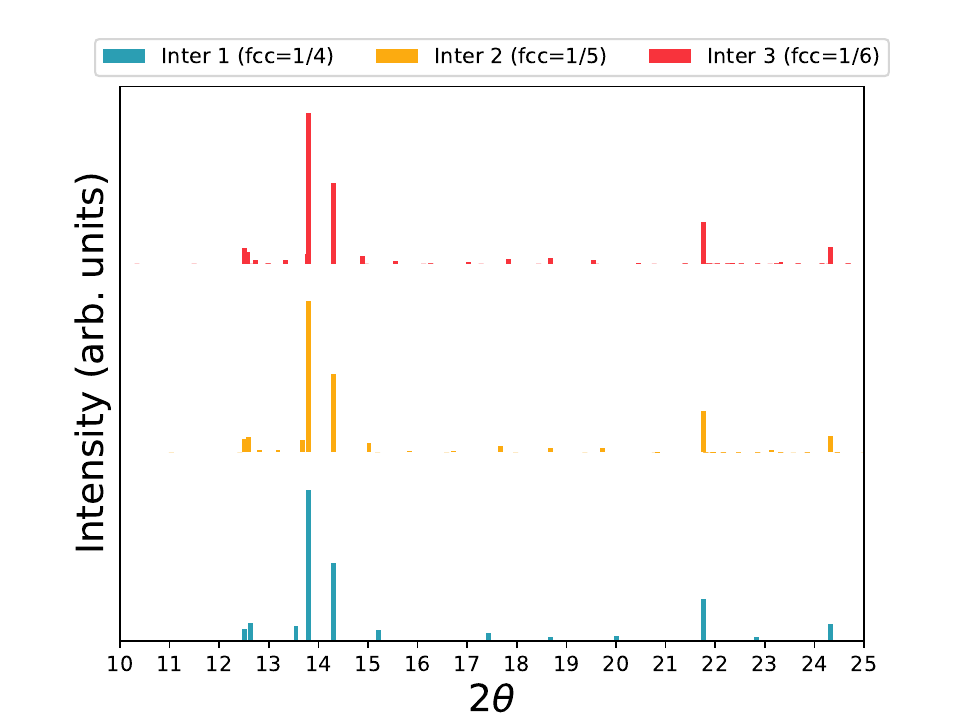}
  \caption{Predicted X-ray powder diffraction pattern for theoretical structures with ratios of $fcc$ layers between $hcp$ and Sm-type.}
  \label{fig:a_xrd_inter}
\end{figure}

\begin{figure}[!h]
  \centering
  \includegraphics[width=\linewidth]{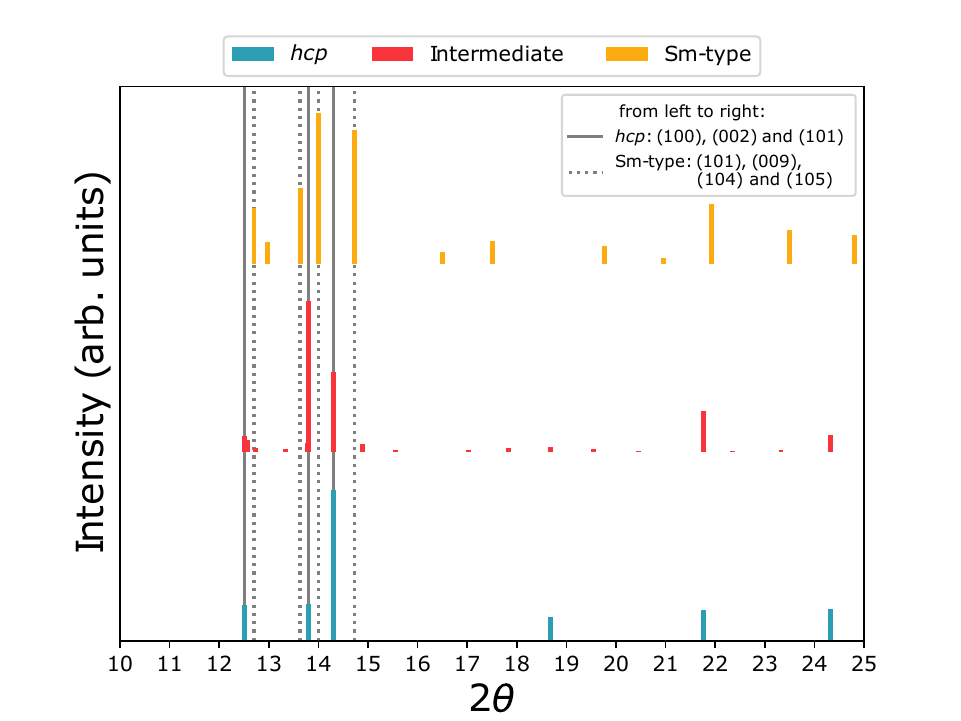}
  \caption{Comparison between predicted X-ray powder diffraction patterns for $hcp$, $Sm$-type and a theoretical intermediate structures of gadolinium. The intermediate structure is designed to have a ratio of $hcp$ (5/6) and $fcc$ (1/6) layers in the middle of $hcp$ and $Sm$-type (2/3 $fcc$) structures. Main low-angle reflections of the $hcp$ ($Sm$-type) structure are represented by vertical solid (dotted) planes.} 
  \label{fig:a_xrd_all}
\end{figure}

The patterns from the different structures can be compared in Fig.~\ref{fig:a_xrd_inter} and in Fig.~\ref{fig:a_xrd_all} we compare the intermediate structure with a $fcc$ ratio of 1/6 with the patterns from the $hcp$ and $Sm$-type structures.

Comparing the patterns of the three intermediate phases in Fig.~\ref{fig:a_xrd_inter}, we observe only slight differences in the major peaks. Mostly, we observe the appearance and splitting of small peaks from the major peaks at $2\theta$ around 12.6 and 13.8. The extent of these splittings increases with the ratio of $fcc$ layers, and is consistent with the splitting of the (100) and (002) reflections of the $hcp$ structure in the XRPD pattern of the $Sm$-type structure. However, these differences should be too subtle to be observed experimentally since the peak broadening would conceal the differences. As a result, we only need to compare one of the proposed structures with the $hcp$ and Sm-type structures without loss of generality.

In Fig.~\ref{fig:a_xrd_all} becomes evident that the major peaks of the intermediate phase share features to the patterns of both $hcp$ and $Sm$-type, not existing any major peak that could exclusively be attributed to the intermediate phase. The broadening of the peaks measured in experiments would make the task even more difficult, since the relative intensities become less clear, as is evidenced in the patterns obtained in Ref.~\cite{Mito2021RelationshipPressures}. Moreover, peak width increases in experiments when more than two phases are present. Therefore, distinguishing any combination of $hcp$ and $Sm$-type phases with the intermediate phase becomes virtually impossible.

In light of these observations, we conclude that previously reported XRPD patterns are not conclusive on the formation and accumulation of $fcc$ stacking faults in structural transformations under pressure.

\section{Enthalpy}
\label{app:enthaply}
The assumption of a linear formation of stacking faults along pressure was tested with the enthalpy curves calculated for the $hcp$ and $Sm$-type structures. To accomplish this, we computed energy-volume curves for both structures using the PBE \cite{Perdew1996GeneralizedSimple} exchange-correlation potential (motivated by a better performance in describing the structural properties \cite{Locht2016StandardApproximation}). Afterwards, we fit the Murnaghan equation of state \cite{Murnaghan1944} to obtain the enthalpy values. Analyzing the calculated enthalpy, we identified the pressure at which the $Sm$-type structure becomes stable to be approximately 3.34 GPa, close to the experimental values.

\begin{figure}[!h]
  \centering
  \includegraphics{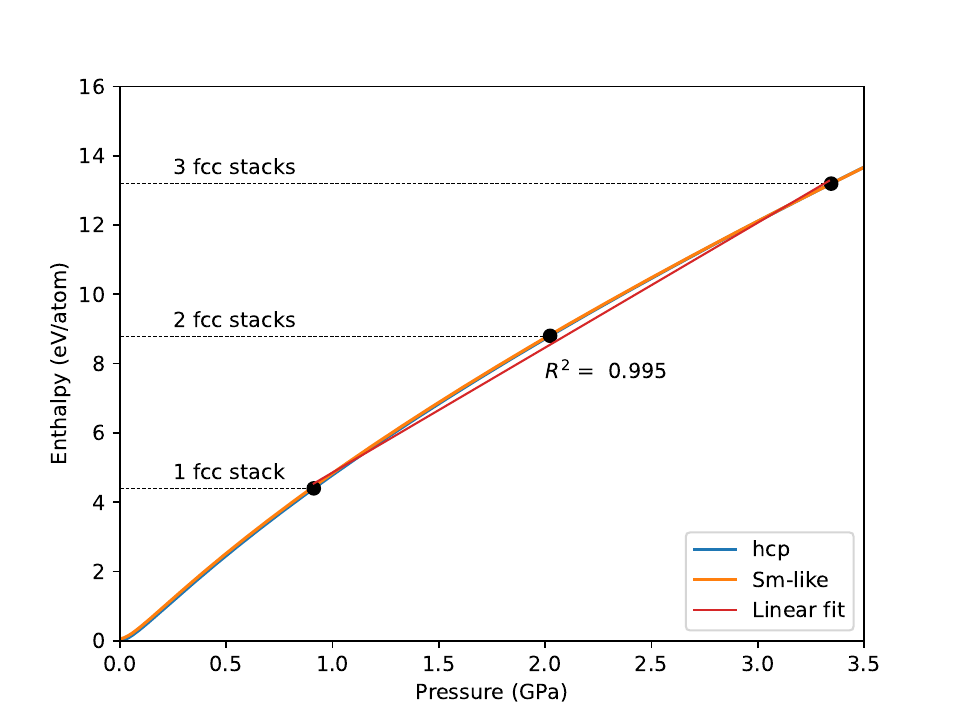}
  \caption{Verification of linear assumption on the formation of stacking faults between $hcp \rightarrow Sm$-like phases from respective enthalpy curves.}
  \label{fig:stacks}
\end{figure}

Under the hypothesis that the structural transition is driven by the accumulation of $fcc$ stacking faults, we associated the enthalpy at this pressure point with the energy required to form three stacking faults. Assuming a constant energy cost for creating a single stacking fault (leading to the linear relation), we pinpointed the enthalpy values on the $hcp$ curve, which correspond to the formation of the first and second $fcc$ layers. We then performed a linear regression on the identified points, as depicted in Fig.~\ref{fig:stacks}. Based on the results of this regression and the corresponding coefficient of determination, we can reasonably conclude that the assumption of a linear progression of stacking faults with increasing pressure is a reasonable approximation.

\section{Magnetic configuration of the dhcp structure}
\label{app:dhcp}
\begin{figure}[!h]
    \centering
    \includegraphics[width=0.5\linewidth]{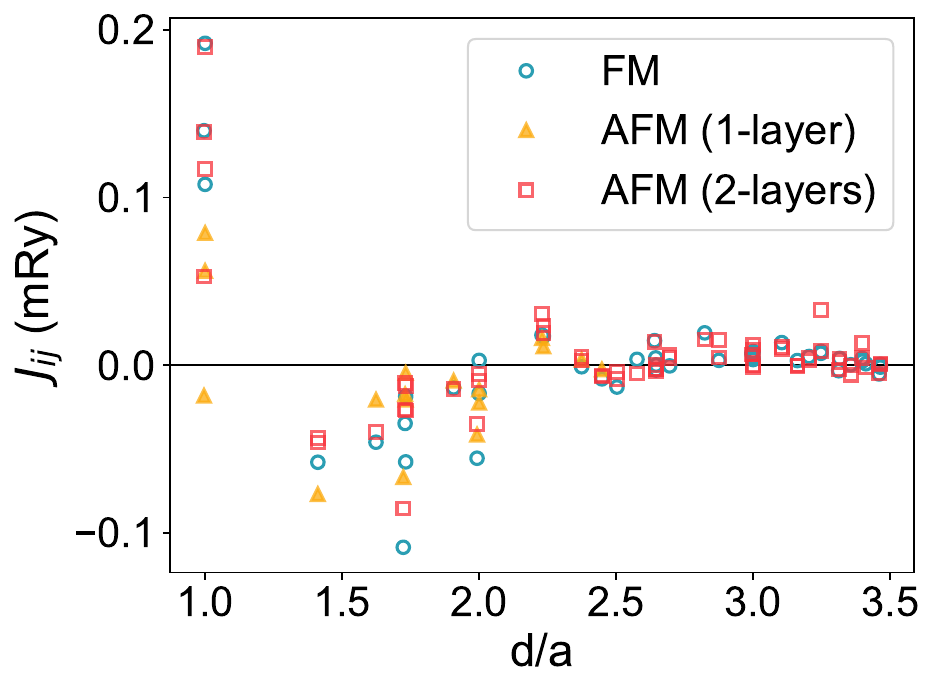}
    \caption{Exchange parameters for Gd in the $dhcp$ structure obtained under different magnetic configurations.}
    \label{fig:a_Jij_dhcp}
\end{figure}

\begin{figure}[!h]
    \centering
    \includegraphics[width=0.75\linewidth]{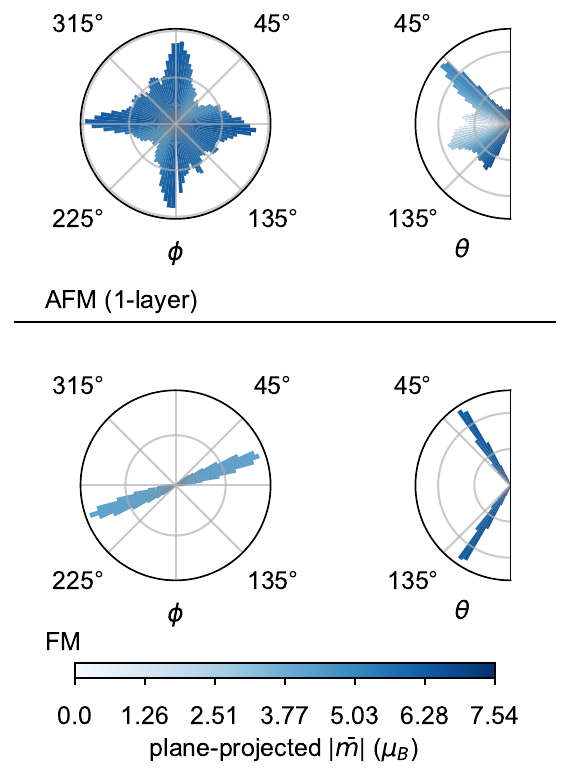}
    \caption{Statistical distribution of spin directions (spherical coordinates with azimuthal angle $\phi$ in the x-y plane and polar angle $\theta$) for different Gd systems simulated T=1K with UppASD. The bars are coloured according to the average magnetic moment (projected in the respective x-y or z plane) of the spins aligned in the respective direction.}
    \label{fig_a_Mag_dhcp}
\end{figure}

The exchange parameters for the $dhcp$ calculated for a sample of magnetic configurations are presented in Figure~\ref{fig:a_Jij_dhcp}. The different sets present a similar behaviour with slight variations in the magnitude of the couplings, with AFM (1-layer, $\uparrow \downarrow \uparrow \downarrow$)  configurations deviating more from the remaining. Due to the strong competition between AFM and FM couplings, the small deviations result in very different magnetic configurations in the Monte Carlo simulations as seen in Figure~\ref{fig_a_Mag_dhcp}.

\section{Heat capacity}
\label{app:heat_capacity}

\begin{figure}[!h]
\centering
   \includegraphics[width=0.75\linewidth]{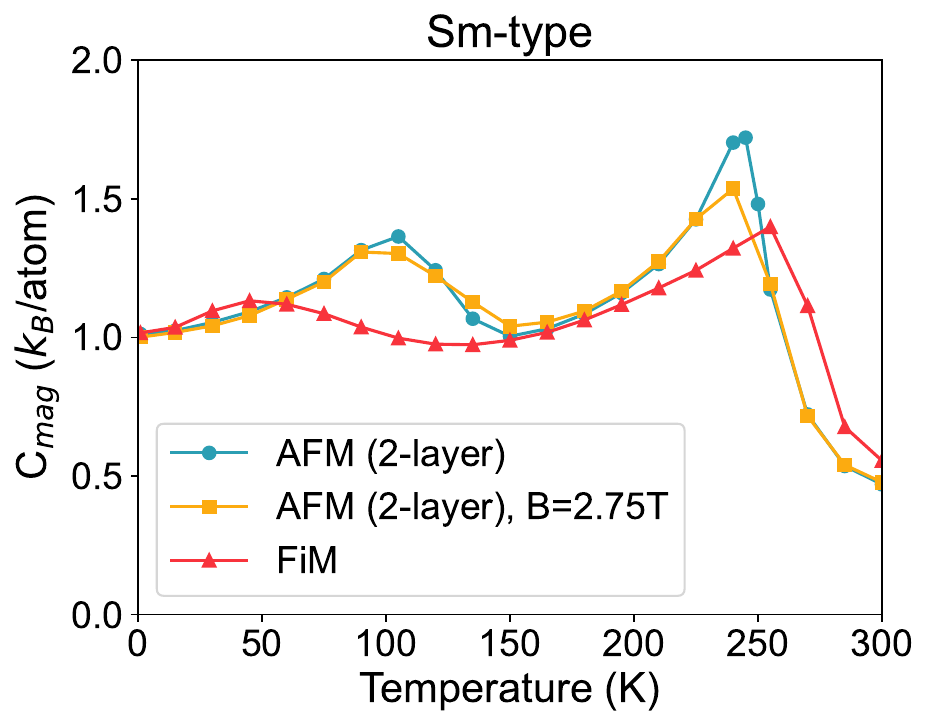}
   \caption{Magnetic heat capacity and computed from Monte Carlo simulations for Gd in the Sm-type structure. }
   \label{fig:MC_Sm}
\end{figure}

Figure~\ref{fig:MC_Sm} shows the temperature-dependent magnetic heat capacity ($C$) calculated in Monte Carlo simulations with UppASD. To calculate the heat capacity, we used the thermodynamic relation:
\begin{equation}
C = \frac{\partial \langle E \rangle}{\partial T}
\end{equation}
where E is the energy evaluated from the Heisenberg Hamiltonian (see Equation~\ref{eq:Heis}).
The peaks in the calculated heat capacity are used as a reference for the magnetic ordering temperature.

\section{Linear Fit}
\label{app:linear_fit}
\begin{figure}[!h]
    \centering
    \includegraphics[width=0.75\linewidth]{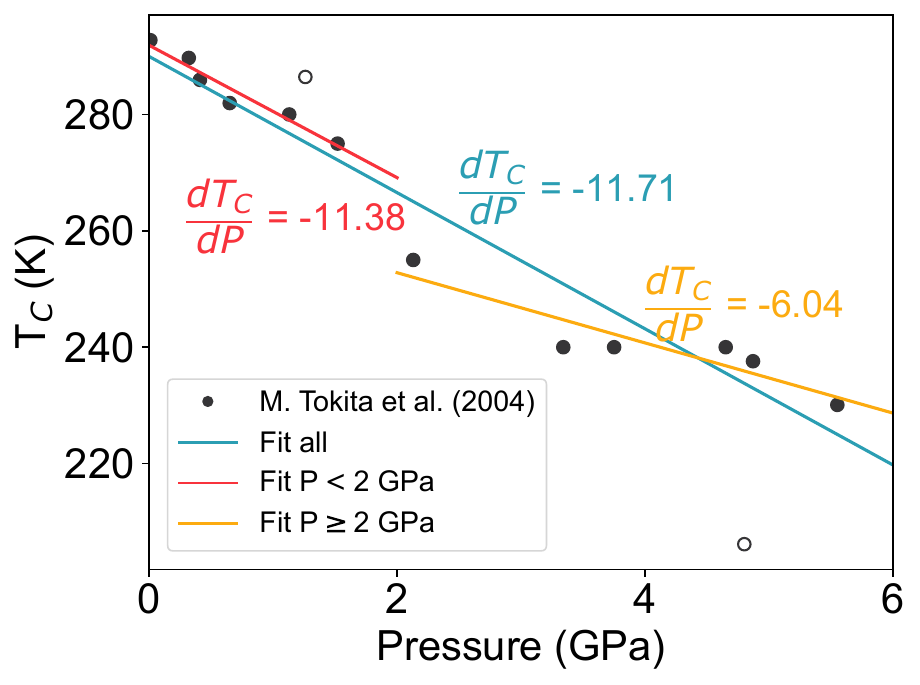}
    \caption{Comparison of linear fits for different pressure intervals. The resultant $dT_C/dP$ varies significantly if a different fit is made for the $hcp$ and $Sm$-type phases. Experimental data from \cite{Tokita2004RKKYRegions}}
    \label{fig:fit}
\end{figure}

In Figure~\ref{fig:fit}, we compare the $dT_C/dP$ ratio obtained for different linear fits of recent experimental data~\cite{Tokita2004RKKYRegions}.  Two cases are considered: when all data points in the 0-6 GPa data range are included,  and the separate fits for the 0-2 GPa and 2-6 GPa ranges.  We observe that the $dT_C/dP$ obtained by fitting solely to the 2-6 GPa is considerably smaller (in magnitude) than the analogous calculations in the 0-2 GPa.  Consequently, by fitting all the data points simultaneously, the change in the trend is missed, and the resulting $dT_C/dP$ follows the trend of the 0-2GPa, where this ratio is sharper. 

The different $dT_C/P$ imply that the magnetic properties vary differently according to the implied structures, highlighting the magnetism of Gd has structure intrinsic properties.

\FloatBarrier
\bibliography{main}
\end{document}